\documentclass[twocolumn]{webofc}
\usepackage[varg]{txfonts}   

\begin{document}
\title{Superheavy Nuclei in the Quark-Meson-Coupling Model}

\author{\firstname{Jirina} \lastname{Stone}\inst{1}\fnsep\thanks{\email{jirina.stone@physics.ox.ac.uk}} \and
        \firstname{Pierre} \lastname{Guichon}\inst{2}\fnsep\thanks{\email{pierre.guichon@cea.fr}} \and
        \firstname{Anthony} \lastname{Thomas}\inst{3}\fnsep\thanks{\email{anthony.thomas@adelaide.edu.au}}
}

\institute{Department of Physics, University of Oxford, OX1 3PU United Kingdom, \\
  Department of Physics and Astronomy, University of Tennessee, TN 37996 USA
\and
           CEA/IRFU/SPhN  Saclay, F91191 France
\and
           CSSM and CoEPP, Department of Physics, University of Adelaide, SA 5005 Australia
          }

\abstract{%
  We present a selection of the first results obtained in a comprehensive calculation of ground state properties of even-even superheavy nuclei in the region of  96 < Z < 136 and 118 < N < 320 from the Quark-Meson-Coupling model (QMC).  Ground state binding energies, the neutron and proton number dependence of quadrupole deformations  and Q$_\alpha$ values are reported for even-even nuclei with 100 < Z < 136  and compared with available experimental data and predictions of macro-microscopic models.  Predictions of properties of nuclei, including Q$_\alpha$ values, relevant for planning future experiments are presented. 
}
\maketitle
\section{Introduction}
\label{sec-1}
One of the main difficulties in solving the nuclear many-body problem is to incorporate nuclear medium effects into the calculation. In contrast to, for example, elementary quantum electrodynamics, where the force between two electric charges is quantified by the Coulomb law and forces among many electrons can be calculated precisely using the principle of superposition, no equivalent approach exists in nuclear physics. Scattering experiments with free nucleons provide information about bare N-N potentials, but an involved numerical treatment is necessary to transform these potentials to a form which desribes the forces between hadrons in the hadronic environment. There have been many theoretical attempts to overcome this problem but they all suffer from the uncertainty associated with many empirical parameters which are not uniquely constrained, thus preventing reliable predictions of nuclear properties in regions not accessible to experiment. One particular area of interest concerns the transuranic superheavy nuclei, where one may hope to discover new chemical elements, not existing in nature but synthesized in the laboratory. Theoretical predictions guiding such experiments are of major interest, especially information about spherical and deformed shell closures and related islands of stability. Given the ambiguity in models based on nucleonic degrees of freedom, alternative approaches have been sought.

The idea of modeling the effective nuclear interaction in-medium using quark degrees of freedom originated in the late 1980s with work by Guichon and collaborators \cite{guichon1988,guichon1996, guichon2004}. They suggested that the origin of nuclear many body forces and of the saturation of nuclear forces can be found in the modification of the structure of a nucleon when it is imbedded in a medium consisting of other nucleons. The model, referred to as the Quark-Meson-Coupling (QMC) model, has been applied to finite nuclei \cite{guichon2006,stone2016}, nuclear matter and neutron stars (see e.g. \cite{stone2007,whittenbury2014} as well as hypernuclei \cite{guichon2008} and a variety of other problems~\cite{Saito:2005rv}. The  model is based on the idea that, instead of the usual treatment of modeling nuclear forces through exchange of mesons coupled to nucleons, taken as point-like particles, this exchange takes place directly between quarks in different nucleons, taken as a cluster of confined valence quarks having a structure in the form, for example, of the MIT bag. When the quarks in one nucleon interact \textit{self-consistently} with the quarks in the surrounding nucleons by exchanging a $\sigma$ meson, the effective mass $M^*_N$ of the nucleon  is no longer linear in the scalar mean field ($\sigma$). It is expressed as
\begin{equation} 
{\rm M^*_{\rm N}=M_{\rm N} - g_{\sigma N}\sigma +(d/2) (g_{\sigma N} \sigma) ^2 },
\end{equation}
 where $g_{\sigma N}$, the $\sigma$ nucleon coupling constant in free space, is a parameter of the model.  By analogy with electromagnetic polarizabilities, the coefficient $d$, calculated in terms of the nucleon internal structure, is known as the “scalar polarizability” \cite{guichon1988}. The appearance of this term in the nucleon effective mass is sufficient to lead to nuclear saturation. This demonstrates a clear link between the internal structure of the nucleon and fundamental properties of atomic nuclei.
\begin{table}
\centering
\caption{Neutron numbers corresponding to proton and neutron drip lines, derived from the Fermi energy for isotopes of elements 96 < Z <136 }
\label{tab-1}
\begin{tabular}{|ccc|ccc|} \hline
Z  &  N(p)  &  N(n)  &  Z  &  N(p)  &  N(n)  \\ \hline
96   &  132  &  224  &  118  &  174  &  278 \\
98   &  134  &  226  &  120  &  180  &  286 \\ 
100 &  138  &  230  &  122  &  184  &  290 \\
102 &  138  &  236  &  124  &  188  & 296 \\  
104 &  146  &  240  &  126  &  192  & 298 \\                                
106 &  146  &  242  &  128  &  196  & 302 \\
108 &  154  &  246  &  130  &  202  & 306 \\
110  &  158 &  250  &  132  &  208  & 310  \\
112  &  164 &  256  &  134  &  214  & 314\\
114  &  168 &  260  &  136  &  218  & 314 \\
116  &  170 &  268  &         &          &        \\   \hline
\end{tabular}
\end{table}

Models of nuclei in the region above Z = 100 can be divided into two classes \cite{sobiczewski2007}, mean field models with effective density dependent interactions in the Hartree-Fock approximation \cite{bender2001,bender2003,bender2013,heenen2015,dobaczewski2015,afanasjev2003,prassa2013}, and macro-microscopic models based on the liquid-drop model combined with Strutinsky-type shell corrections (see e.g. \cite{moller2016,moller1997,muntian2003,muntian2003a}. Although these models yield similar results for some bulk properties of superheavy nuclei, they differ in important details. Superheavy nuclei exist mainly because of their shell structure, in turn, dependent on level densities close to the Fermi level. It is therefore very important to understand their single-particle spectrum, in particular the location and magnitude of proton and neutron shell gaps. In addition, it may be expected that the competition between the Coulomb repulsion and the surface tension in many nucleon systems would play an important role and the sharp shell closures, familiar in lighter nuclei, may be diluted to form regions of a distinct spherical or deformed nature around a particular (N,Z) shell closure.

The various models predict different proton and neutron shell closures  above $^{\rm 208}$Pb. The macroscopic–microscopic models yield a `canonical' island of stability at Z = 114, N = 184 \cite{bender2001,sobiczewski2007}, while the self-consistent mean-field models (SC) non-relativistic models predict  Z = 126, N = 172 - 184 \cite{bender2001,bender2003} and the relativistic mean-field models (RF) favour Z = 120, N = 172 \cite{bender2001,afanasjev2003}. The most recent comparison of traditional models applied to superheavy nuclei may be found in \cite{heenen2015} but no way to reconsile these differences is offered.

It is therefore interesting to examine the predictions of the QMC model which belongs to the class of mean field models but differs in key ways from both non-relativistic models with the Skyrme or Gogny interaction and classical relativistic field models in important details \cite{stone2016}.  In this paper we present a selection of the first results obtained in a comprehensive study of the ground state properties of spherical and axially symmetrical even-even superheavy nuclei in the region  96 < Z < 136 and 118 < N < 320.  A new version of the QMC model, QMC$\pi$, is used in this work, where the published model~\cite{stone2016} has been extended to include pion-exchange.  The calculational procedure is described in section~\ref{sec-2} and the resulting ground state binding energies of selected isotopes and the evolution of the deformation parameter $\beta_{\rm 2}$ for bound isotopes of elements with 100 < Z < 128 and isotones with 148 < N < 204, along with the Q$_\alpha$ values are presented  in section~\ref{sec-3}. Section~\ref{sec-4} contains predictions of the QMC$\pi$ model for properties of superheavy nuclei considered in planning future experiments, followed by section~\ref{sec-5} summarizing the current results and outlining future work.
\begin{figure}
\centering
\includegraphics[width=7.5cm,clip]{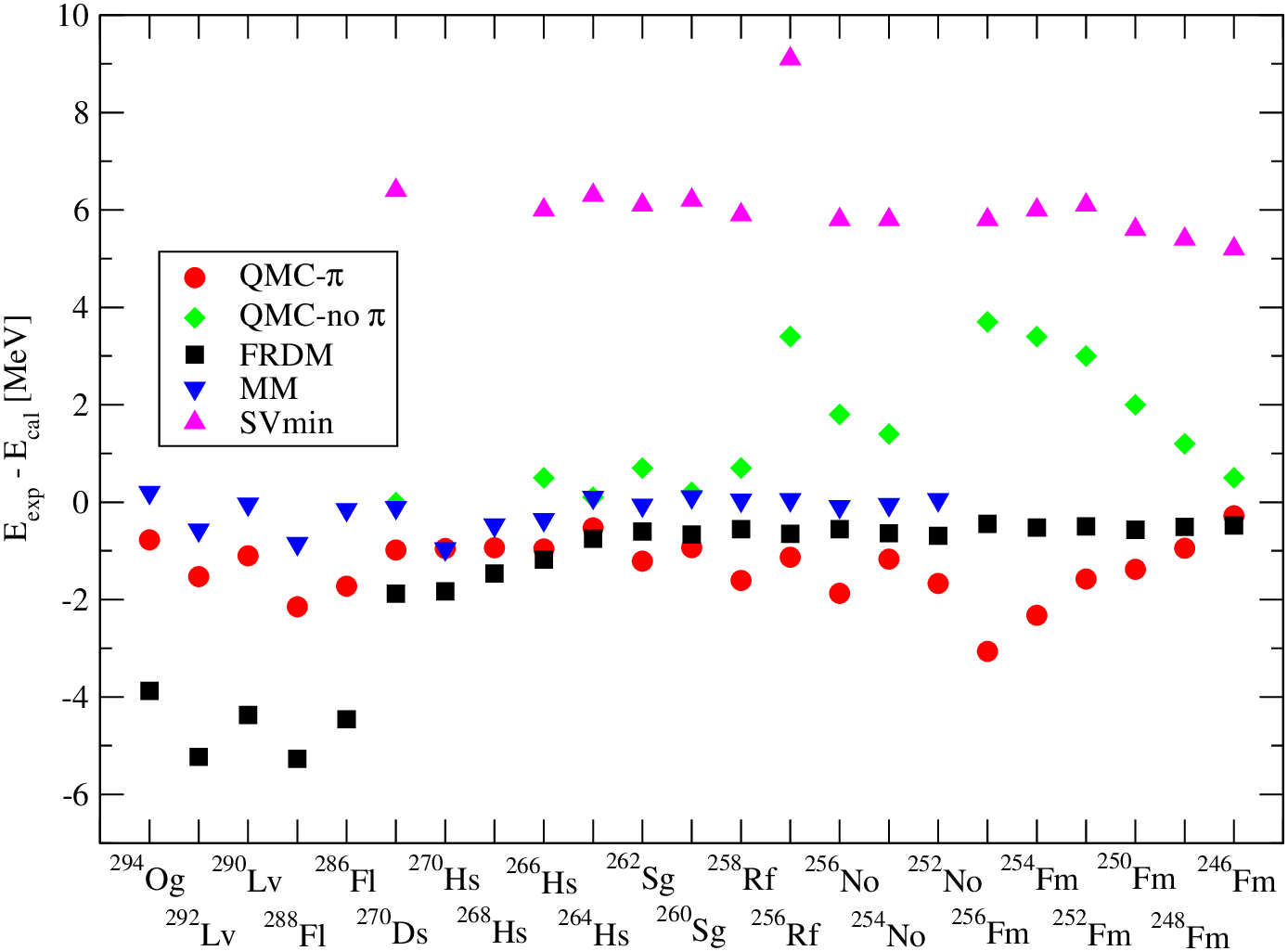}
\caption{(Color online). Ground state binding energies of selected 'benchmark' even-even superheavy nuclei. The experimental data were taken from \cite{huang2017,wang2017}. }
\label{fig-1}
\end{figure}

\section{Method of calculation}
\label{sec-2}
The QMC$\pi$ energy density functional (EDF) has been constructed in the same way as described in \cite{stone2016}, except that the explicit pion exchange has been included in the Hamiltonian. We used HF+ BCS code SKYAX allowing for axially symmetric and reflection-asymmetric shapes, adapted by P.-G. Reinhard \cite{reinhard2016}  for use with the QMC EDF. The best parameter set was sought using the experimental data set by Kl\"{u}pfel et al. \cite{klupfel2009} and the fitting package POUNDERS \cite{kortelainen2010,wild2015}.  The volume pairing in the BCS approximation as been adopted with proton and neutron pairing strength fitted to data in \cite{klupfel2009}. We note that the addition of the explicit pion exchange in the model  did not increase the number of  parameters beyond the four used in the previous work, namely G$_\sigma$, G$_\omega$, G$_\rho$ and M$_\sigma$, but its addition was reflected in slight changes (less than 5\%)  from the values reported in \cite{stone2016}. The new parameter set is compatible with nuclear matter properties E$_0$  =-15.8 MeV, $\rho_{\rm 0}$ = 0.153 fm $^{-3}$, K$_{\rm 0}$ = 319 MeV, S$_{\rm 0}$ = 30 MeV and L = 27 MeV. 
\begin{figure}
\centering
\includegraphics[width=7.5cm,clip]{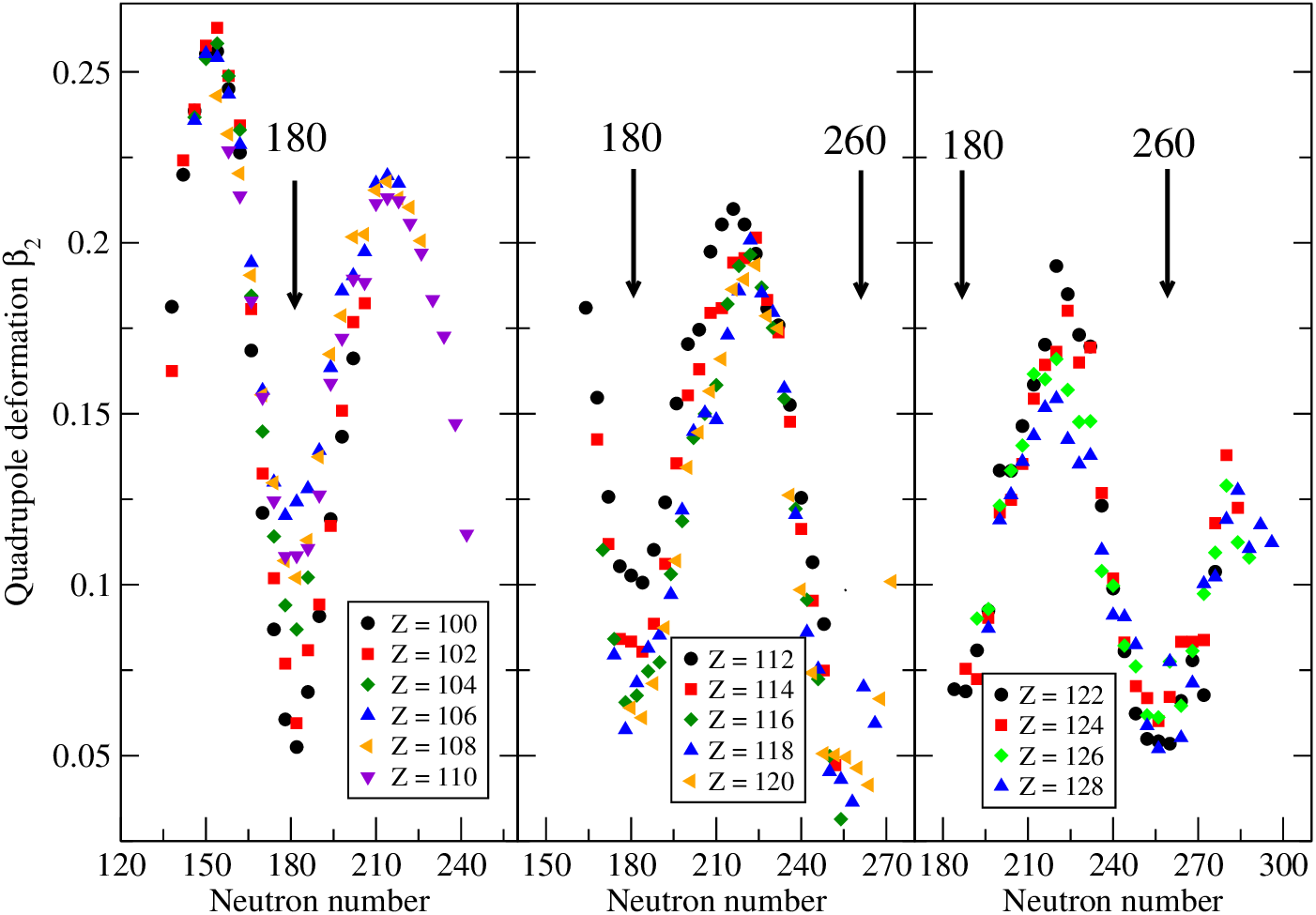}
\caption{(Color online). Quadrupole deformation calculated in QMC$\pi$ for isotopes with proton number 100 $<$ Z $<$ 128.}
\label{fig-2}
\end{figure}

\begin{figure}
\centering
\includegraphics[width=7.5cm,clip]{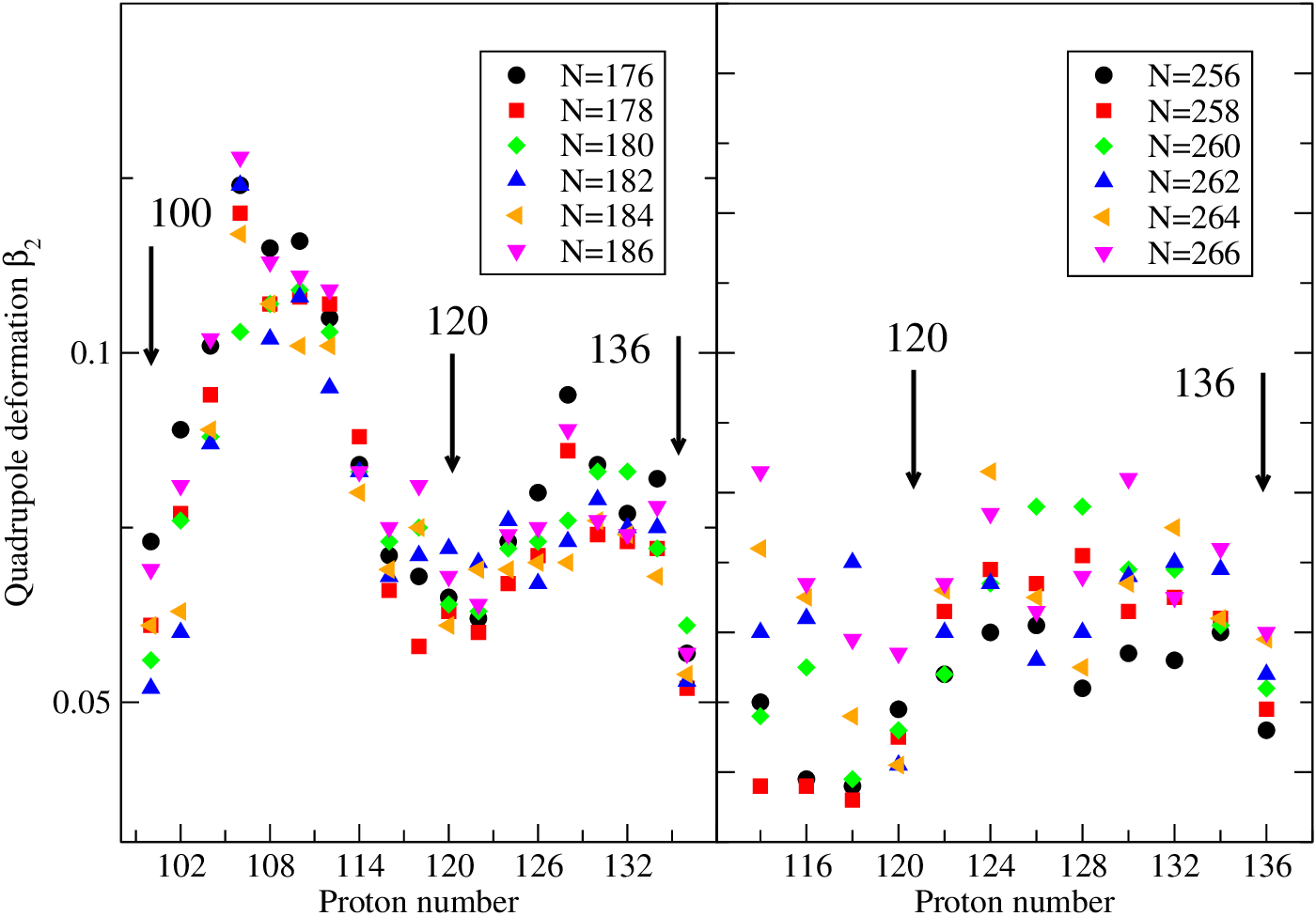}
\caption{(Color online). The same as figure~\ref{fig-2} but for isotones with neutron number 176$<$ N $<$ 186 (left panel) and 256$<$ N $<$ 266 (right panel).}
\label{fig-3}
\end{figure}
The selection of isotopes studied was limited by the requirement that only bound single particle states are occupied; that is, the Fermi level does not increase to the continuum.  This condition limited the range of neutron number for each element as given in Table~\ref{tab-1}.
\section{Ground state properties and decay of selected superheavy nuclei}
\label{sec-3}
\subsection{Binding energies and shapes}
\label{sec-3.1}
Figure~\ref{fig-1} illustrates the difference between experimental and calculated ground state binding energies of selected superheavy nuclei for which the experimental data is known well enough that they may be regarded as 'benchmark' data \cite{dullmann2017} to calibrate theoretical models for subsequent use in uncharted areas of the nuclear landscape, notably in the transuranic region. These results illustrate a rather dramatic change from a significant underbinding, yielded by the mean field model with the Skyrme SVmin parameter set \cite{klupfel2009} and, to a lesser extent, predicted by QMC-without-pion model in the Z=100 - 104 region \cite{stone2016},  to a slight overbinding with the present QMC$\pi$, microcroscopic-macroscopic models Finite-Range-Droplet-Model (FRDM) \cite{moller2016} and of Muntian et al. (MM) \cite{muntian2003,muntian2003a}. It is interesting to observe that the inclusion of pion-exchange in the QMC model has a significant effect on the ground state binding energies.
\begin{figure}
\centering
\includegraphics[width=7.5cm,clip]{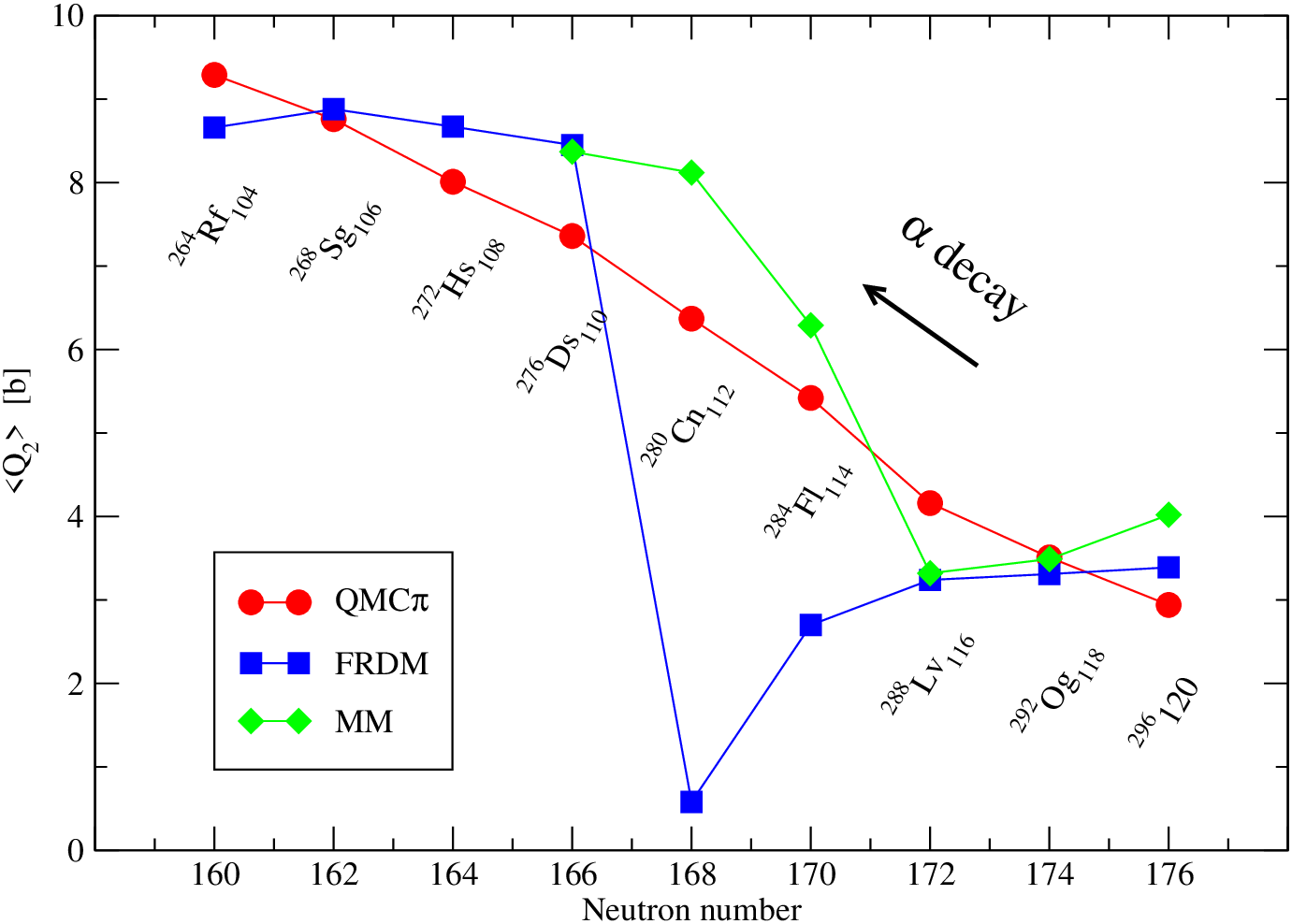}
\caption{(Color online).  Intrinsic quadrupole moments of even-even nuclei in the $\alpha$-decay chain of $^{\rm 296}$120, as calculated in the QMC$\pi$, FRDM and MM models. See Fig.~1 in \cite{shi2014} and Ref.~\cite{dullmann2017a} for a comparison with other models and the text for more explanation.}
\label{fig-4}
\end{figure}
As mentioned already in the introduction, one of the main motivations for modeling the superheavy region is to locate regions of relative stability. These regions would be the best targets for synthesis of either new elements or unknown isotopes for elements already known. The deformation parameters, $\beta_{\rm 2}$, determined in the QMC$\pi$ model by minimizing the ground state binding energies, are plotted with respect to neutron (proton) number in Figs.~\ref{fig-2} and \ref{fig-3}, respectively. The results in Fig.~\ref{fig-2} clearly identify two regions of low deformation. Values of  $\beta_{\rm 2}$ less than about 0.1 are predicted for isotopes of all elements with 100 < Z < 128  with neutron number in the region around N$\sim$ 180  and, for Z > 114 also in the vicinity of  N = 260. The calculation of isotones with 176 < N < 186 reveals a strong indication for low deformation close to  Z = 100, 120 and 136 (or possibly higher), as illustrated in Fig.~\ref{fig-3} (left panel). The isotones with 256 < N < 266  generally exhibit low deformation in the region of 116 < Z < 136 (right panel). The effect is less prominent at Z = 120, but still clearly present at Z = 136. The analysis of the quadrupole deformation in QMC$\pi$ shows no evidence for the Z = 114 shell closure predicted in the FRDM and MM models nor at Z = 126, as suggested in SC models. However, it does agree with the results of RF models finding a shell closure at Z = 120. The neutron shell closures at N = 172, found in the SC and RF models is also not confirmed in this work, while the N=180 region is somewhat close to the value N = 184 found in the FRDM, MM and SC models. The region of sphericity we predict at N = 260 and Z = 136 is outside the reach of current experimental techniques but may potentially lead to a new island of stability. Further calculation, in particular of spectra of single-particle states, is needed to take this problem further.

Next we examine in more detail the development of the quadrupole deformation of isotopes along $\alpha$ decay chains of special interest, as they are a prime tool for identification of new elements (keeping in mind that spontaneous fission is in competition with $\alpha$ decay).  In Fig.~\ref{fig-4} we show the predicted mean intrinsic quadrupole moment calculated in this work. This is related to 
 $\beta_{\rm 2}$ as ${\rm <Q_{\rm 2}>} = {\rm (3/(4\pi) A R_{\rm 0}^{\rm 2} \beta_{\rm 2} }$, with R$_{\rm 0}$=1.2 A$^{\rm 1/3}$. We note that this definition \cite{bender2013} is consistent with the procedure used to calculate the deformation parameters in this work. There are alternative definitions in the literature, used for example in \cite{shi2014}. Therefore, only the qualitative neutron number dependence of ${\rm <Q_{\rm 2}>}$ can be compared between different calculations. We see a remarkably smooth decrease  in deformation with increasing neutron number along the decay chain in the QMC$\pi$ model prediction, similar to that found by Shi et al. \cite{shi2014} in calculations using UNEDF1 and UNDEF1$^{\rm SO}$ Skyrme forces. In contrast, the FDRM microscopic-macroscoscopic model predicts a distinctive spherical shape at N = 168. The MM model more or less shows a decrease in deformation with increasing neutron number with a slight tendency to sphericity at N = 172. It is important to realize that deformation is treated differently in mean-field and micro-macro models (for details see \cite{heenen2015}). Thus experimental data on quadrupole moments of superheavy elements may provide important information, leading to some distinction between the two classes of model to be used in the superheavy region. The recent progress in laser resonance ionization spectroscopy \cite{laatiaoui2016,block2017} offers a prospect that electromagnetic moments, charge radii and isotope shifts of superheavy isotopes will be accessible experimentally in the near future.
\begin{table}
\centering
\caption{Q$_{\alpha}$ in MeV calculated in the QMC$\pi$, FRDM and MM models. Experimental data without errors have been taken from nuclear mass tables \cite{wang2012}, while entries with errors are from original papers \cite{oganessian2007,oganessian2015,dvorak2006,nishio2010}.}
\label{tab-2}
\begin{tabular}{|cccrrr|}
\hline
Z  &  N  &  Q$_{\alpha}^{\rm exp}$ &  Q$_{\alpha}^{\rm QMC\pi}$ &  Q$_{\alpha}^{\rm FRDM}$ & Q$_{\alpha}^{\rm MM}$ \\ \hline
102  &  150  &   8.548 &   8.20 &   8.82 &   8.53 \\
102  &  152  &   8.226 &   8.10  &  7.97 &   8.06 \\
102  &  154  &   8.581 &   8.00  &  8.57 &   8.36  \\
104  &  152  &   8.926 &   9.40  &  8.75 &   8.93 \\
104  &  154  &   9.190 &   9.10  &  9.32 &   9.29  \\
106  &  154  &   9.901 &   10.10 &  9.93 &   9.95  \\
106  &  156  &   9.600 &   10.00  &  9.61 &   9.49   \\
108  &  156  &   10.591 &  11.00 &  10.57 &  10.59 \\
108  &  158  &   10.346 &  10.60 &  9.69   &  10.04  \\
108  &  160  &   9.62$\pm$ 0.16 & 10.30 &  9.00 & 9.49 \\
108  &  162  &   9.02$\pm$ 0.03 & 10.20 &  8.69 &  8.78 \\
110 &   160  &  11.117                 & 11.10 &  10.30 & 11.36  \\
112 &   172 &   10.230                 &  10.40 & 8.69 &  9.76  \\
114  &  172 &   10.33$\pm$0.08  & 11.30 &  9.39 &  9.76 \\
114  &  174 &   10.08$\pm$0.06  & 11.00 &  9.16 &  10.32  \\
116 &   174 &   11.00$\pm$0.08  &  11.60 & 11.12 & 11.08  \\
116 &   176 &   10.80$\pm$0.07  &  11.40 & 10.82 & 11.06 \\
118 &   176 &   11.81$\pm$0.06  &  12.20 & 12.28 & 12.11 \\  \hline
\end{tabular}
\end{table}
\subsection{Q$_\alpha$ values}
\label{sec-3.2}
Knowledge of $\alpha$ decay life-times is crucial to  predicting properties of the $\alpha$ decay chains of superheavy elements, which, as pointed out above, are crucial in the search for and detection of new elements and their isotopes. The $\alpha$-decay life-times are exponential functions of the energy release, Q$_\alpha$,  in the decay, which, in turn, depends on the mass difference between the parent and daughter states. This means that while the absolute values of the nuclear masses are not crucial in this context, the differences are essential. We adopt here the expression for the $\alpha$-decay half-life given in Ref.~\cite{moller1997} for an even-even parent with the atomic number Z
\begin{equation}
{\rm log (T_{\rm 1/2}/s)} = {\rm (aZ + b)(Q_\alpha /MeV)(cZ + d)}
\end{equation}
with a=1.66175, b=-8.5166, c=-0.20228 and d=-33.9069. It follows, for example, that a difference in Q$_\alpha$ of the order of 1 MeV in a nucleus with Z = 118 would make a difference in T$_{\rm 1/2}$ of three orders of magnitude. Thus the calculation of Q$_\alpha$ as close to reality as possible is vital for planning experiments.
\begin{figure}
\centering
\includegraphics[width=7.5cm,clip]{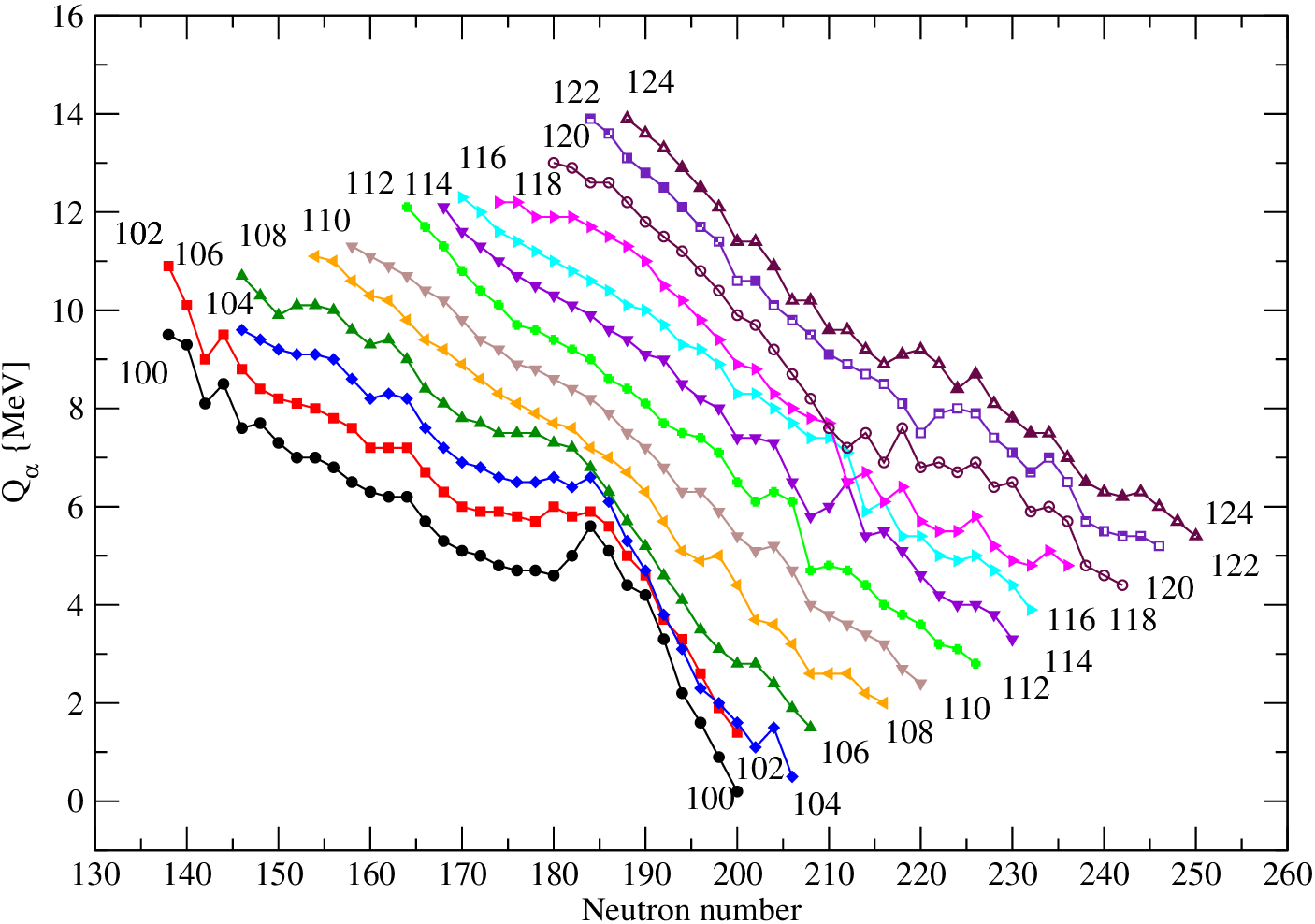}
\caption{(Color online). Values of the $\alpha$ particle separation energy, Q$_\alpha$, calculated in QMC$\pi$ for isotopes with 100 < Z < 120 in the region of  neutron numbers 138 < N < 252.}
\label{fig-5}
\end{figure}
\begin{figure}
\centering
\includegraphics[width=7.5cm,clip]{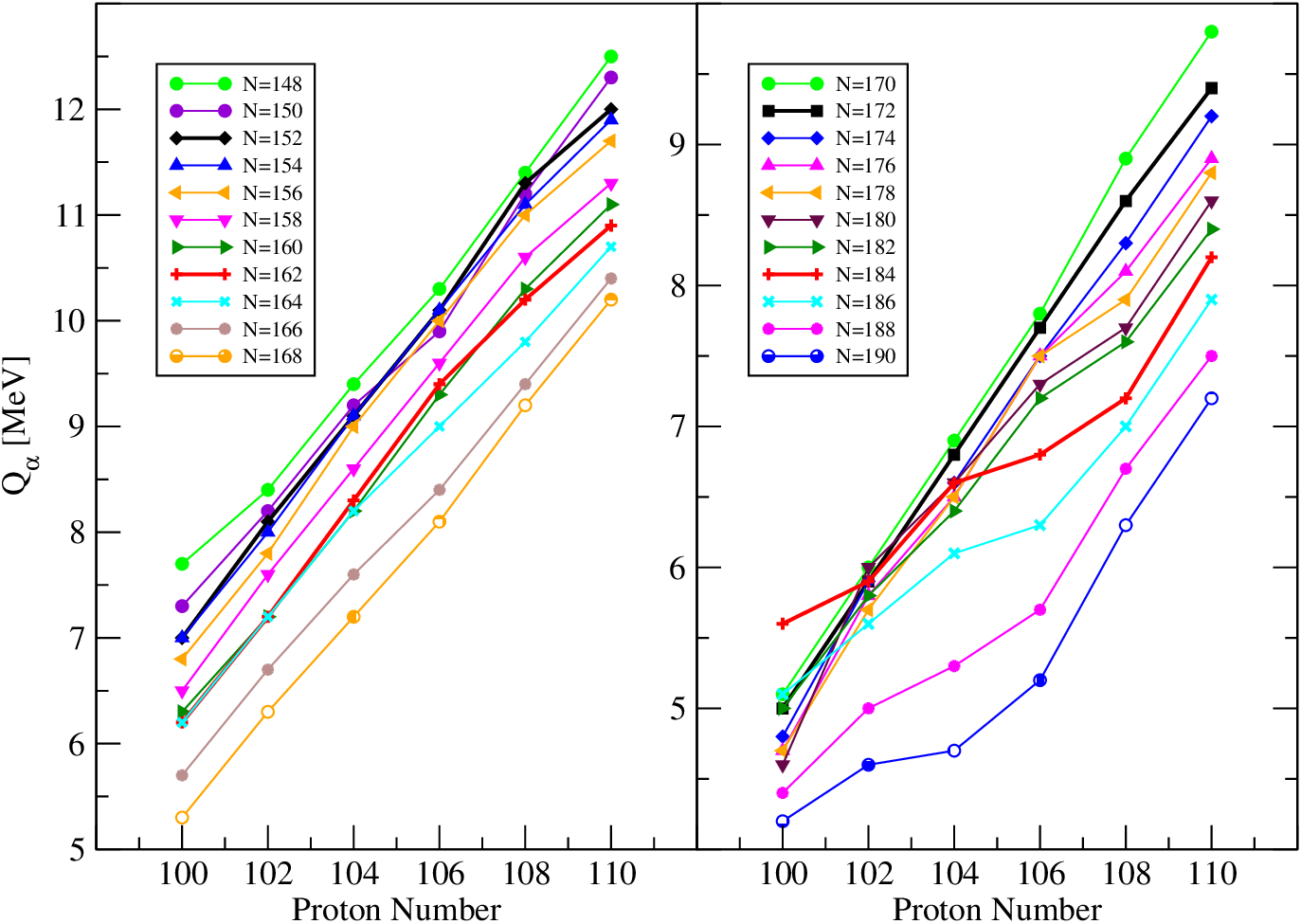}
\caption{(Color online). The same as Fig.~\ref{fig-5}  but for isotones with 148 < N < 190  in the region of proton numbers 100 < Z < 110. The N=152, 162, 172 and 184 systems, corresponding to the suggested shell closures in different models, are highlighted by thicker lines in both panels.}
\label{fig-6}
\end{figure}
The neutron number dependence of Q$_\alpha$ obtained in this work is illustrated in Fig.~\ref{fig-5}. It is beyond the scope of this work to compare the QMC$\pi$ predictions for Q$_\alpha$ with all the models in the literature. We refer the reader to, for example, \cite{heenen2015,carlsson2016,jachimowicz2014,wang2016} and references therein. The important result, which has not been observed in any of the other models, is that the (weak) effects of the N = 152 and 162 shell closures disappear in nuclei with Z > 108, while the effects are enhanced for  N$\sim$180. We predict a smooth neutron number dependence of  Q$_\alpha$ for N<200 for all elements with Z up to 124, not showing any effects of shell structure. Some variations may be indicated for higher N but at this stage we can draw no systematic conclusion. To our knowledge such behaviour of Q$_\alpha$ in this region has not been predicted before and a follow up would be very interesting.
\begin{table*}[!ht]
\centering
\caption{Predicted properties of selected isotopes of Rf, Hs, Fl, Og, El-120 and  El-124, as calculated in the QMC$\pi$, FRDM and MM models. Experimental binding energies are taken from \cite{huang2017,wang2017}. All energies are in MeV. }
\label{tab-3}
\begin{tabular}{|cccl|rrr|rrr|rrr|}
\hline
Isotope            & Z   &  N & E$_{\rm bin}^{\rm exp}$  & E$_{\rm bin}$ & $\beta_{\rm 2}$ & Q$_\alpha$ &E$_{\rm bin}$ & $\beta_{\rm 2}$ & Q$_\alpha$ &E$_{\rm bin}$ & $\beta_{\rm 2} $& Q$_\alpha$  \\ \hline
            &     &      &                    &  \multicolumn{3}{|c|}{QMC$\pi$} & \multicolumn{3}{|c|}{FRDM} & \multicolumn{3}{|c|}{MM}   \\ \hline
$^{\rm 252}$Rf   &104&148&             & 1860.20  &0.245 & 9.4   &1860.29 & 0.251   &  9.54  &    1859.29          & 0.245 &	9.85         \\
$^{\rm 268}$Sg & 106 & 162&1963.37  &1965.30 &  0.229&  9.4  & 1965.39   & 0.232  &  7.59 &    1964.48 &	0.233  &7.89 \\
$^{\rm 262}$Hs  &108&154&             & 1911.80  &0.228 & 11.1 &1912.32 & 0.243   & 10.88 &   1911.21    &          0.244   &  11.00  \\
$^{\rm 272}$Hs  &108&164&1981.79 & 1983.80  &0.236 & 9.8   &1984.48 & 0.220  & 9.20   &   1982.98     &          0.225  &   9.80        \\
$^{\rm 272}$Ds   & 110&  162 & 1973.36 & 1974.2 & 0.238 & 10.90  & 1975.56  &  0.221 &  10.04 &  1973.90 &  0.226 &  10.74 \\
$^{\rm 284}$Fl   &114&170&2034.01 & 2035.40  &0.129 & 11.6 &2037.98  & 0.064  & 9.44   & 2033.85 & 0.149   & 11.53  \\
$^{\rm 296}$Og &118&178&             & 2095.80  &0.095 & 11.9 &2099.12  &-0.063  & 12.29 & 2094.67 &0.039    & 12.06  \\
$^{\rm 296}$120&120&176&             & 2082.90  &0.065 & 13.2 &2085.70  & 0.075  & 13.69 & 2081.83 & 0.085   & 13.23  \\
$^{\rm 298}$120&120&178&             & 2097.30  &0.063 & 13.0 &2100.16  & 0.040  & 13.35 & 2095.48 & 0.054   & 13.44 \\
$^{\rm 306}$124&124&182&             & 2124.80  &0.076 & 15.0 &2126.39  & 0.000  & 13.43 &              &             &           \\   \hline
\end{tabular}
\end{table*}

Values of Q$_\alpha$ in the region Z$\leq$110 are given in Fig.~\ref{fig-6}, showing the proton number dependence of Q$_\alpha$ for 148 < N < 190. We see a relatively smooth increase in Q$_\alpha$ with increasing Z for N $\leq$ 168, with a minor sensitivity to N = 152 and N = 162 (highlighted by thicker lines in the left panel of Fig.~\ref{fig-6}. This pattern prevails to N = 170 -174 (top part of the right panel) before a notable development of lower Q$_\alpha$ values, varying systematically with Z (bottom part of the right panel).

The outcome of the QMC$\pi$ model indicates that there is a subtle interplay between proton and neutron degrees of freedom in developing regions of nuclei with increased $\alpha$-decay half life. As already discussed in the introduction and in the literature (e.g. \cite{bender2001}), it is likely that the sharp shell closures and shape changes observed in lighter nuclei, will instead be manifest as smoother patterns around the expected ``shell closures''. These patterns have their origin in the competition between the Coulomb repulsion and surface tension of the large nuclear systems in which the single-particle structure is only one of the critical ingredients. 
\section{Future experiments}
\label{sec-4}
Following discussions during the Fusion2017 meeting \cite{dullmann2017,dullmann2016} we provide predictions for binding energies, quadrupole deformations and Q$_\alpha$ values for selected nuclei summarized in Table~\ref{tab-3}. Corresponding results from the FRDM and MM models have been added for comparison. The reason for interest in these nuclei comes from their calculated shell structure in some traditional models, for example $^{\rm 260}$Hs (N = 152, Z = 108 both supposed to be deformed shell closures). The presence of the N = 162 shell closure should manifest itself in  $^{\rm 268}$Sg and hence influence the Q$_\alpha$ value of  $^{\rm 272}$Hs. The Z = 108 shell closure could be studied by examining the properties of  $^{\rm 272}$Ds. The existence of an 
$\alpha$-branch from the decay of  $^{\rm 272}$Ds is expected but has not yet been seen, the likelihood depending on the Q$_\alpha$ value. The prospect of observing  $^{\rm 296}$Og,  $^{\rm 296,298}$120 and, possibly,  $^{\rm 306}$124, is considered to be realistic using available beams and targets.

Although all three models in Table~\ref{tab-3}  agree reasonably well in their predictions of binding energies and deformations, the differences in  Q$_\alpha$ are significant (see section~\ref{sec-3.2}). The most striking result is that QMC$\pi$ gives a much higher value of Q$_\alpha$ than those yielded by the FRDM and MM models. For example, the calculated $\alpha$-decay half-live of  $^{\rm 268}$Sg in the QMC$\pi$, FRDM and MM models  are 0.21 s, 3.6 d and 5.9 h, respectively.  One of the reasons for these variations is the presence of the N=162 shell closure in the models. The higher  Q$_\alpha$ value is consistent with existence of the shell closure. This result also supports the suggestion of a higher Q$_\alpha$ for  $^{\rm 272}$Hs. The calculated values of Q$_\alpha$ for $^{\rm 284}$Fl, $^{\rm 296}$Og,  $^{\rm 296,298}$120  and $^{\rm 306}$124 lead to very short T$_{\rm 1/2}$, 1.4x10$^{\rm -5}$ s, 4.0x10$^{\rm -4}$ s,  2.3x10$^{\rm -6}$s, 5.8x10$^{\rm -6}$ s and 1.0x10$^{\rm -8}$ s, respectively. More model predictions in this regions can be found in \cite{dullmann2017a}.
\section{Summary and conclusions}
\label{sec-5}
We have presented the first results of the application of the QMC$\pi$ model to superheavy nuclei. Having in mind that the model is dependent on only four, well constrained variable parameters with a clear physical meaning, it is encouraging to observe that  predictions for ground state binding energies, axially symmetrical shapes, regions of shell closure and Q$_\alpha$ values are in good agreement with experimental data where available. The results presented in this work are of a very similar quality as the outcomes of other models, mean-field and macroscopic-microscopic which depend on many more parameters. Of course, the most interesting observation is that some of the predictions of the QMC$\pi$ model differ significantly  from those in the traditional models and only experimental data can decide which model is closest to reality.   Naturally there are some limitations in this work.  As the energy surfaces as a function of quadrupole deformation may be complicated in the superheavy region (see e.g. \cite{heenen2015}), it would be desirable to confirm the present results for nuclear shapes and the corresponding binding energies by repeating the minimization of the total binding energy with a constraint on the quadrupole moment. This procedure would help to avoid local minima and make a decision in cases of coexisting shapes with very similar binding energies. In addition, the predicted regions of shell closures, deduced in the present work from the neutron and proton number dependence of quadrupole deformation may be made more precise by examination of the spectrum of proton and neutron energy levels. These extensions of the current calculation are in progress.
\section{Acknowledgement}
It is a pleasure to thank Chris D\"{u}llmann for suggestions and comments, concerning experimental data, their significance and prospects. The contribution of Kay Marie Martinez, who performed the fitting of the QMC$\pi$ parameters using the POUNDERS package, is gratefully acknowledged. This work was supported by the University of Adelaide and the Australian Research Council through the ARC Centre of Excellence for Particle Physics at the Terascale and grant DP150103101.

\end{document}